%% file: main.tex
\documentclass[journal]{vgtc}                     

\onlineid{1100}

\vgtccategory{Research}

\vgtcpapertype{system}

\title{\system: A Visual Analytics Solution for \\Automated Speech Recognition Analysis and Exploration}

\author{Sunwoo Ha, Chaehun Lim, R. Jordan Crouser, and Alvitta Ottley} %

\authorfooter{
\item
 S. Ha, C. Lim, and A. Ottley are with Washington University. E-mails: sha@wustl.edu, chaelim@wustl.edu, alvitta@wustl.edu.
 \item
 R. J. Crouser is with Smith College. \\E-mail: jcrouser@smith.edu.
}

\usepackage{stackengine}
\usepackage{tabularx}
\usepackage{amsfonts}
\usepackage{amsmath}
\usepackage{bbm}
\usepackage{subfig}
\usepackage[dvipsnames]{xcolor}
\usepackage{colortbl}
\usepackage[normalem]{ulem}
\usepackage{enumitem}
\usepackage[hidelinks]{hyperref}
\usepackage{url}
\newcommand{\system}[0]{\textbf{\textsc{Confides}}}

\newcommand{\linkicon}[0]{\includegraphics[width=.012 \textwidth]{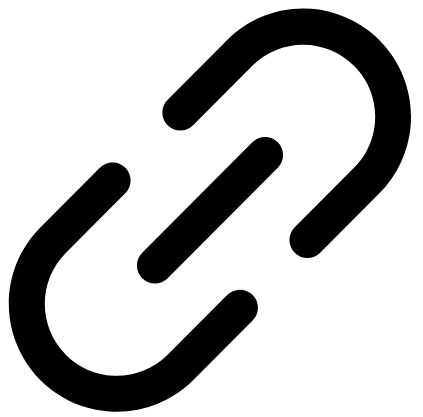}}

\teaser{
  \centering
  \includegraphics[width=0.93\linewidth]{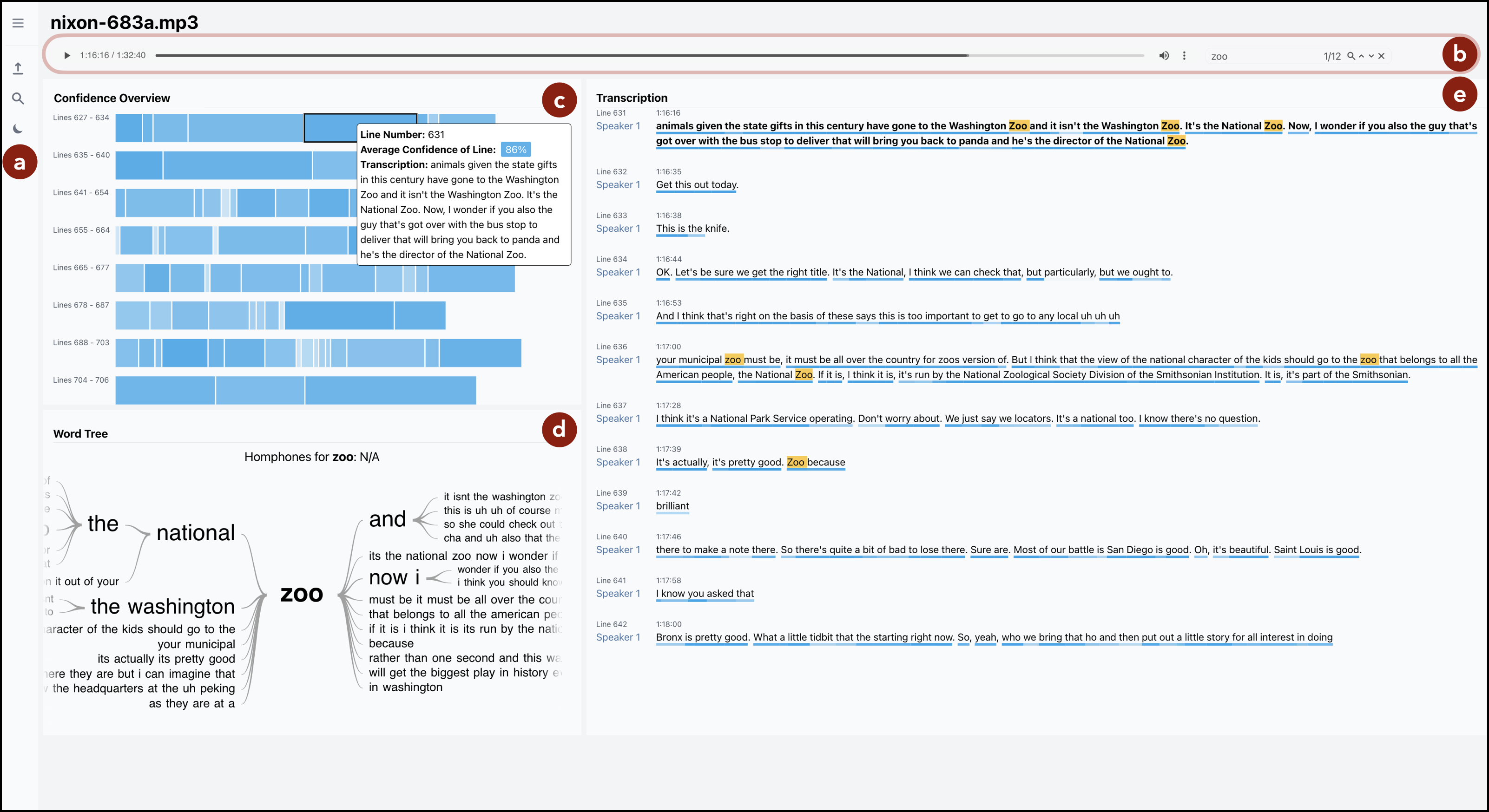}
  \caption{An overview of \system: (a) The collapsible side menu contains controls for login, selecting, uploading, and transcribing audio files via AWS Transcribe. (b) At the top of the dashboard are the audio player and search bar. 
  (c) The confidence overview displays the length and average confidence value of each line segment in the transcription (encoded by the width and opacity of each rectangle, respectively). (d) The word tree provides context to a specific search term and shows which words most often follow or precede it. (e) The user can view and edit the transcription; each word is underlined, and its opacity indicates the confidence score. 
}
  \label{fig:teaser}
}

\abstract{
  Confidence scores of automatic speech recognition (ASR) outputs are often inadequately communicated, preventing its seamless integration into analytical workflows. In this paper, we introduce \system, a visual analytic system developed in collaboration with intelligence analysts to address this issue. \system\ aims to aid exploration and post-AI-transcription editing by visually representing the confidence associated with the transcription. 
  We demonstrate how our tool can assist intelligence analysts who use ASR outputs in their analytical and exploratory tasks and how it can help mitigate misinterpretation of crucial information. We also discuss opportunities for improving textual data cleaning and model transparency for human-machine collaboration.
} 


\begin{document}




\maketitle

\input{content/content}

\acknowledgments{
We are grateful for the feedback provided by Syrine Matoussi, Lan Kung, Christine Brugh, and our LAS collaborators. This work is supported by the National Science Foundation under Grant Nos. 2118201 and 2142977.
It is also based upon work done, in whole or in part, in coordination with the Department of Defense (DoD). Any opinions, findings, conclusions, or recommendations expressed in this material are those of the authors and do not necessarily reflect the views of the DoD and/or any agency or entity of the United States Government.}

\bibliographystyle{abbrv-doi}

\bibliography{main}
\end{document}

%% file: content/content.tex
\newpage
\section{Introduction}

The amount of audio data grows exponentially with each passing hour.
However, turning raw audio data into actionable intelligence can be complex, as it relies heavily on the accuracy of the algorithms, transcription services employed, and the quality of the original audio files. Despite significant advances in Artificial Intelligence (AI) and Automatic Speech Recognition (ASR), the analysis of audio data remains fraught with inaccuracies and is notably time-consuming~\cite{berke_deaf_2017}. As a result, fields such as intelligence analysis urgently require solutions that enable them to efficiently and accurately harness this vast and rapidly expanding resource.

Existing solutions that provide automatic speech-to-text and editing capabilities of the transcription (i.e., Rev, Happy Scribe) fall short as they often inadequately or rarely highlight and convey metrics that may bring forward potential inaccuracies to the user. The lack of transparency between the machine and the analyst within these existing services hinders the complete integration and effective utilization of speech-to-text technology within their workflows and analytical processes~\cite{sacha_role_2016}. Relying on AI, especially in human-machine collaborations, without awareness of these uncertainties can be detrimental, as the quality of an analyst's work is a direct result of the trust~\cite{dasgupta_familiarity_2017} in and accuracy of the information presented to them.

To address this challenge, we present a new visual analytics system called \system, which we developed in collaboration with intelligence analysts. Our system enhances understanding of speech-to-text results by showing how confident the model is in its transcription. It uses the speech-to-text service from Amazon Web Services (AWS) and aims to make exploring and editing post-AI transcription easier by providing visual representations of confidence levels. By using interactive visualizations and multiple views, we demonstrate how our tool could improve the analysis of speech-to-text output and foster trust in human-machine collaborations. We also discuss how our system could support common analytical and exploratory tasks for intelligence analysts working with audio transcriptions. Finally, we explore opportunities to improve model transparency and textual data cleaning to encourage more effective human-machine collaboration. Through this work, our main contributions include:
\begin{itemize}
    [noitemsep,topsep=0pt]
    \setlength{\parskip}{0pt}
    \setlength{\parsep}{0pt} 
    \item We outline specific design goals derived from tasks that intelligence analysts perform within their workflow when inspecting audio transcriptions.
    \item We develop a visual analytics system, \textbf{Confides}, that allows analysts to easily transcribe their audio files and explore the transcription while being aware of uncertainties within the data.
    \item To demonstrate the applicability and usefulness of the system, we present a realistic use case scenario that forages for information from the Nixon White House tapes.
\end{itemize}

\section{Related Work}
We briefly overview the prior work on transparency in human-machine collaborations in visual analytics and AI-assisted decision-making settings and visualizing confidence in speech-to-text outputs.

\subsection{Communicating Model Uncertainty}
From the point of view of mixed-initiative systems, Sacha et al.~\cite{sacha_role_2016} discussed the role of uncertainty, awareness, and trust in visual analytics and argued that users’ confidence in the machine teammate's results depends on their degree of awareness of the different types of uncertainty that are present or generated in the system. Additionally, research has shown that visually communicating uncertainty can support users’ interpretation of data and support decision-making ~\cite{fernandes_uncertainty_2018, jung_displayed_2015, kay_when_2016}. Therefore, communicating uncertainty is vital, especially when an AI agent assists in a decision-making process as it can potentially mitigate unwanted behaviors from the user such as underutilization~\cite{monadjemi2022guided, dietvorst_algorithm_2015, kim_when_2023, dabek_grammar-based_2017, ha2024guided} or overreliance~\cite{bucinca_trust_2021, jacobs_how_2021, ha2024guided} on AI suggestions. Wang et al.~\cite{wang_are_2021} explored various explanation methods for recidivism prediction and forest cover prediction tasks. Their study found supportive evidence suggesting that providing information about feature contribution allowed participants to be aware of the uncertainties within the model and appropriately calibrate their trust. 

\subsection{Visualizing Confidence in Speech-To-Text Outputs}
Various methods change the appearance of text to convey speech-to-text confidence (e.g., alternating the font size~\cite{piquard-kipffer_qualitative_2015}, font color~\cite{brodlie_review_2012}, font opacity~\cite{berke_deaf_2017}, and underlining~\cite{vertanen_benefits_2008}). Vertanen et al.~\cite{vertanen_benefits_2008} utilized underlining, color, and opacity to visually embed confidence in speech recognition outputs, where the opacity of the colored underline of the text is based on the degree of confidence that the predicted word is accurate. In addition to embedding confidence in the transcription text, Wu et al.~\cite{wu_interactive_2020} developed a prototype visualization system that provides an overview of the confidence of speech-to-text outputs with a bar chart. In this visualization, each bar is mapped to a segment in the transcription, and all the bars of one transcribed text are sequentially visualized. We take inspiration from these prior works to communicate confidence to analysts and design the interactive visualizations used in our system.

\section{Design Goals}
\label{sec:user_tasks}

Numerous applications exist for text-to-speech data, including legal environments where depositions are recorded~\cite{prasad_automatic_2002}, healthcare settings where doctors record audio notes for medical records~\cite{latif_speech_2021}, automatic captioning for online videos, and academic research involving interview studies. Each use case scenario has different and unique tasks. Therefore, to create an ecologically valid solution, we ground our work in government intelligence analysis settings where analysts often analyze transcribed audio data for matters related to national security. 

We partnered with the Laboratory of Analytical Sciences, which facilitates collaborations between the Department of Defense's (DoD) intelligence community and academia. Our tasks and design goals were initially derived from a one-hour, semi-structured interview with a language specialist from the DoD. We inquired about current tools, tasks, and goals that are supported, typical analysis workflows, collaborative activities, pain points, and features that are missing. This initial interview provided context and grounded the design process. We then adopted an iterative design methodology with bi-weekly feedback meetings from January 2023 to December 2023. The design goals presented in this section are the culmination of feedback received from our collaborators from these meetings. Attendees included academic partners, DoD-affiliated analysts, and the projects' program manager. Below we summarize the design goals (\textbf{DG}) for the visual analytics system:
\begin{itemize}
[noitemsep,topsep=0pt]
    \setlength{\parskip}{0pt}
    \setlength{\parsep}{0pt} 
    
    \item \textbf{DG1: Interactively visualize transcription and assess the quality of the transcription output}. Current transcription services lack transparency and rarely communicate the uncertainties and confidence values of automatic transcriptions. In many tasks, especially ones with consequential decisions, the analyst must determine when to trust the automatic transcription versus taking the time to listen to the source audio. Analysts must also determine whether additional data cleaning is necessary. 
    \item \textbf{DG2: Correct and perform basic textual data cleaning}. The system should allow the analyst to easily explore and playback audio segments, and make corrections to the transcription (e.g.,  adding, deleting, viewing alternative words, and replacing the text if desired). 
    \item \textbf{DG3: Discern patterns and extract actionable intelligence within the transcription}. The system should facilitate the extraction of insights and allow the analyst to identify patterns and trends across the audio transcription data. It should enable the analyst to explore the context and frequency of spoken content, enhancing the ability to track discourse over time.
\end{itemize}

To support the analyst in performing these tasks, Confides offers three main interactive views, as detailed in Section~\ref{sec:views}. Additionally, the backend of the system is connected to AWS Transcribe, enabling effortless transcription of audio files and exploration of the outputs in a single interface.

\section{Visual Analytics System Design}
Through several iterative designs and feedback from our collaborators, we developed \system. The following sections describe the framework of the system and how each view will support the analyst with the design goals outlined in Section~\ref{sec:user_tasks}.

\subsection{Framework}
\begin{figure}[!h]
    \centering
    \includegraphics[width=0.9\linewidth]{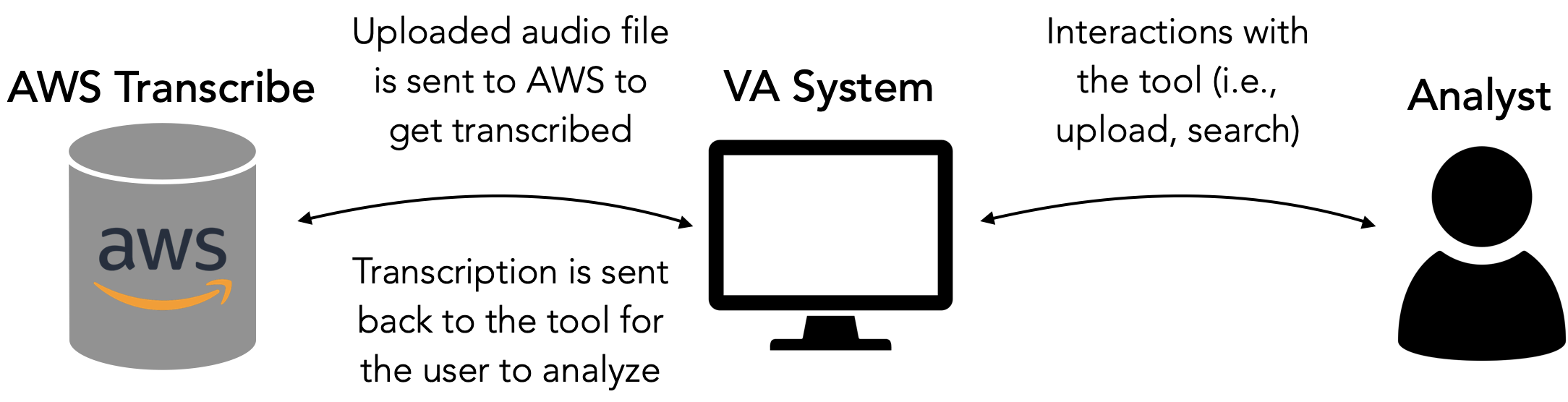}
    \caption{The framework of \system. The audio files are uploaded and sent to AWS for automatic transcription. Users can select which transcriptions to explore and analyze.}
    \label{fig:framework}
    \vspace{-1em}
\end{figure}

The backend of our system utilizes the service offered by AWS Transcribe to automatically transcribe audio files uploaded by the analyst. Despite the vast amount of cloud-based speech-to-text services available, we chose the service provided by Amazon because of its straightforward and well-documented API. Additionally, the transcription output contains valuable insights, such as confidence scores, automatic segmentation, speaker labels, and alternative texts. Once the analyst uploads the audio file with our tool and the transcription of the file is received from AWS Transcribe, the transcription is available for exploration by the analyst. Figure~\ref{fig:framework} shows the framework of the system.

\subsubsection{AWS Transcribe Output}
AWS Transcribe offers a line segmentation feature in their transcription output that will divide the transcript if it detects a pause or change in speaker. The service offers up to a maximum of 10 different speakers and will use its speaker diarization algorithm to determine the speaker at the current time in the audio. AWS Transcribe also supports word-level confidence, which is defined as a value between 0 and 1. Based on their documentation, ``a larger [confidence] value indicates a higher probability that the identified item correctly matches the item spoken"~\cite{AWSTranscribe}. Due to confidentiality, AWS does not release any information on how this confidence score is calculated. Therefore, we stress that the confidence score should not be treated as absolute truth but rather as a metric that offers the analyst a way to gauge the potential flaws within the transcription. 

\subsection{Views}
\label{sec:views}
To promote transparency and encourage calibration of trust in the machine's output, each view of the system is embedded with the machine's confidence in the transcription, whether that is term or segment-based. 

\subsubsection{Confidence Overview}
To provide analysts with a visual overview of the transcription data, we designed the following view shown in Figure~\ref{fig:teaser}(c). In this view, each rectangle element corresponds to a segment in the transcript, with the width representing the audio length of the segment. Similar to the bar chart in the prototype proposed by Wu et al.~\cite{wu_interactive_2020}, the rectangle segments are sequentially ordered from left to right. Each segment's average confidence determines its corresponding rectangle's opacity, meaning the lower the confidence, the more transparent the rectangle will be. This opacity value is calculated by averaging the confidence score associated with every word within the segment. 

Within the confidence overview, analysts can rapidly gauge the confidence of segments and playback certain portions of the audio by clicking on the rectangle element associated with the segment~\textbf{(DG1)}. Clicking on the rectangle element prompts the audio player to move and play the corresponding segment instantly. When the analyst hovers over a rectangle element within the view, a tooltip dynamically updates to provide the analyst with specific details on demand about the corresponding segment. This includes information such as the segment's line number, the rolling average of confidence values within the segment, and the segment's text. 

\subsubsection{Transcription Editor}
The transcription view, as seen in Figure~\ref{fig:teaser}(e), is designed to assist the analyst in all the tasks listed in Section~\ref{sec:user_tasks}. This view displays the automatic transcription of a singular audio file from AWS. Each line represents a segment of the audio spoken by a specific speaker. We utilize color coding to distinguish speakers. 
To promote transparency, we provided both a visual representation (shown by the opacity of the underline -- inspired by Vertanen et al.~\cite{vertanen_benefits_2008}) and a textual representation (the tooltip of each word shows the confidence score) of the confidence for each word within the text (\textbf{DG1}). As the audio plays, the transcription will follow along and automatically scroll to the current segment of the audio being played. The segment currently being played will also be indicated by the boldness of the text within the transcription. By utilizing the search bar shown in Figure~\ref{fig:teaser}(b), the analyst can query for specific keywords and traverse through all instances of the word within the transcription (\textbf{DG3}). The search word is highlighted in yellow within the transcription for easy viewing. The last actionable item the analyst can perform within the transcription view is editing the output. If and when the analyst finds errors within the transcription, the analyst can add or delete text, or even replace a word within the transcription with an alternative suggestion from AWS (\textbf{DG2}).

\subsubsection{Context Word Tree}
Unlike the other two views above, the word tree view will only be populated when the analyst provides a specific keyword they are interested in exploring with the search bar. This view, as seen in Figure~\ref{fig:teaser}(d), depicts multiple parallel sequences of words. Based on visualizations such as the Word Tree~\cite{wattenberg_word_2008} and Sententree~\cite{hu_visualizing_2017}, we visualize a node-link diagram where nodes are words and links indicate word co-occurrence within the same segment. The size of the words in this visualization is proportional to the number of occurrences of the word observed. The average confidence of the keyword is also communicated to the analyst within this view. 

In addition to the visualization, this view provides the analyst with a list of phonetically similar words -- also known as homophones -- to the current search keyword. The goal of presenting homophones in this view is to bring awareness of the nuances of the English language, allow the analyst to discern between similar-sounding words, and reduce misunderstandings in spoken and written discourse. The list of homophones shown to the analyst is gathered from an online platform that has curated an extensive collection of phonetically similar words~\cite{ homophones}.

With the word tree visualization, we aim to show which words most often follow or precede the specific keyword indicated by the analyst (\textbf{DG3}). This view will allow the analyst to explore and understand the context in which the specified word is being said. The analyst can also click on any neighboring word to navigate the tree visualization. 

\section{Case Study: Panda Diplomacy and the Nixon Tapes}

During the Nixon Administration, China gifted two pandas  -- Ling-Ling and Hsing-Hsing -- to the US. This case study will use the \textit{Nixon White House Tapes}~\cite{whitehousetapes} to learn about the \textit{Panda Diplomacy}.
With this case study (developed by Kenney et al~\cite{kenney2022panda}), we demonstrate how analysts can use \system\ to find relevant data and enable them to answer their key intelligence questions more efficiently. We focus on two questions: ``\textit{When were the pandas expected to arrive in the US?}" and ``\textit{Which locations were considered for housing the pandas}?" We provide a video~\footnote{\url{https://youtu.be/hbeDn5D-GCg}} walking through the case study. The reader can also click on the \linkicon\ icons in subsections~\ref{sec:answering} to view the walk-through for that specific question.

\subsection{About the Nixon White House Tapes}
U.S.\ President Richard Nixon's administration secretly recorded conversations held in the White House from 1971 to 1973. These recordings, infamously as the Nixon White House Tapes surfaced during the Watergate Scandal and ultimately led to Nixon's resignation. The administration installed recording devices in the Oval Office and other locations in the White House, intending to document meetings and conversations for historical purposes and to aid in decision-making. As these tapes were captured with concealed microphones, the audio quality is often poor, making the task of transcribing and extracting information after transcription strenuous and overwhelming for analysts. This is the ideal scenario for leveraging the strengths of \system\ and visual analytics. The analysis of this data can reveal pivotal information, but ASR outputs are fraught with inaccuracies due to the poor recording qualities. Since manually transcribing and analyzing the data is tedious, the analyst can offload the computationally heavy work to the machine and utilize their knowledge and perception to generate insights from the transcription output.

\begin{figure}[!h]
    \centering
    \includegraphics[width=\linewidth]{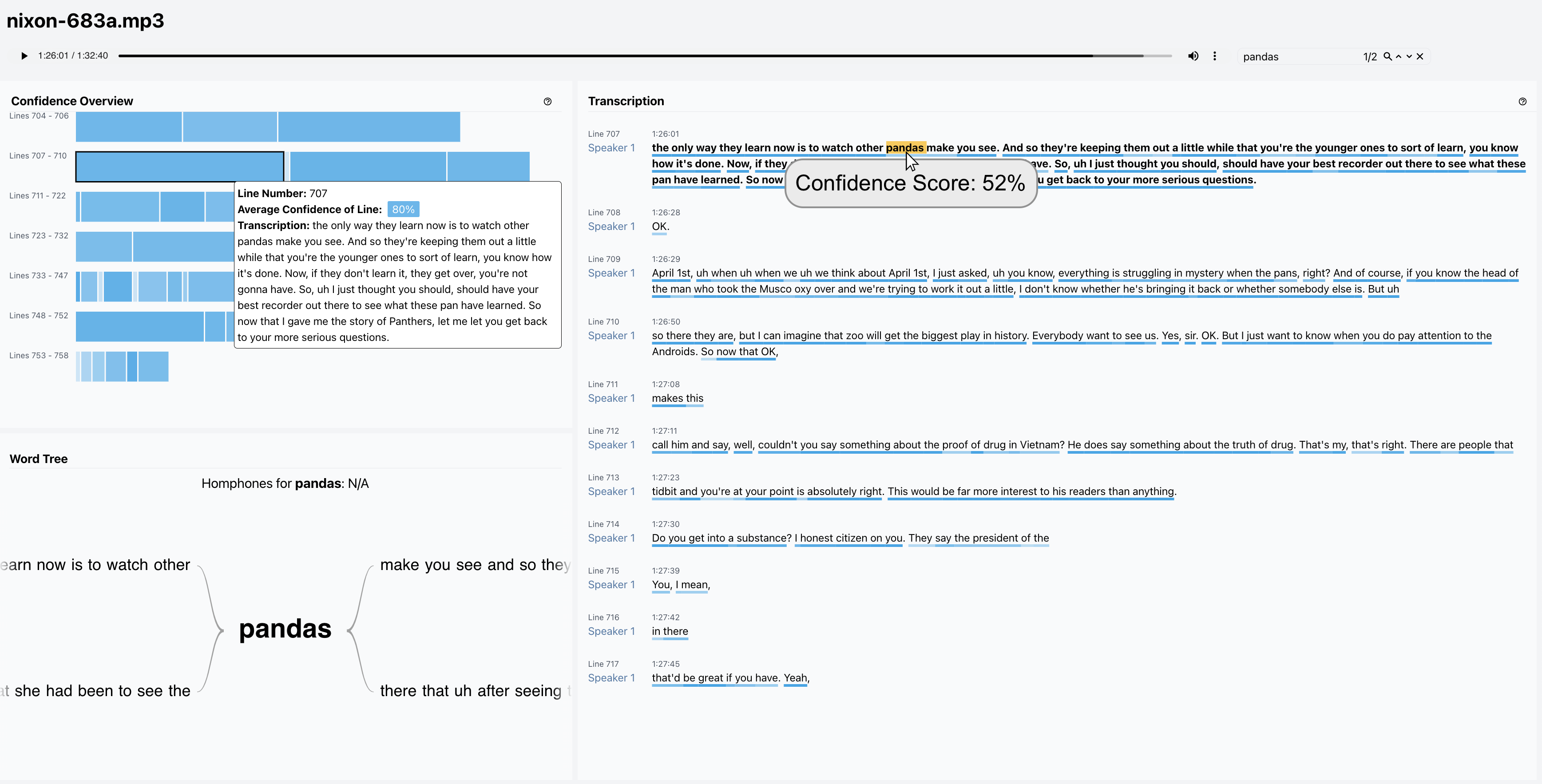}
    \caption{Searching for ``pandas" in the current transcription revealed two instances. We can observe that the first instance of this search term has a confidence score of 52\%.}
    \label{fig:case_study1}
    \vspace{-1em}
\end{figure}

\subsection{Reconnaissance, Quality Assessment, and Editing}
The analyst begins by searching for instances of ``pandas" within the transcription (See Figure~\ref{fig:case_study1}). The query results indicate that ``pandas" is believed to be said only twice within this hour-and-a-half-long audio file. Puzzled, the analyst reads the first segment (line 707) that contains the term ``pandas" and observes that the individual confidence in the term ``pandas" (52\%) is quite low compared to the overall average confidence of this line segment (80\%). However, the terms ``pan" and ``panther" also appear in the segment. Considering that ``pan" is a subword/substring of the desired search term and ``panther" is phonetically similar to ``pandas," the analyst plays the source audio and confirms that ``pandas" is indeed inaccurately labeled as ``pan" or ``panther." The analyst edits the transcription to reflect the audio. We note that there is no search-and-replace feature as a precaution, so the analysts should listen to the audio before making changes to the transcription.

\subsection{Extracting Intelligence and Deciding When to Rely on the AI's Output}
\label{sec:answering}

\paragraph{\hspace{-3mm}\textbf{When were the pandas expected to arrive in the US?} \href{https://youtu.be/hbeDn5D-GCg?si=56DGu--fjOJqB_gn&t=77}{\linkicon}}
While updating the transcription, the analyst discovered that line 709 refers to ``pans'' and ``April 1st,'' the latter with two instances, both with 100\% confidence scores. The analyst decided the source audio was unnecessary and concluded the pandas arrived in the US with high confidence on April 1st.

\begin{figure}[!h]
    \centering
    \includegraphics[width=\linewidth]{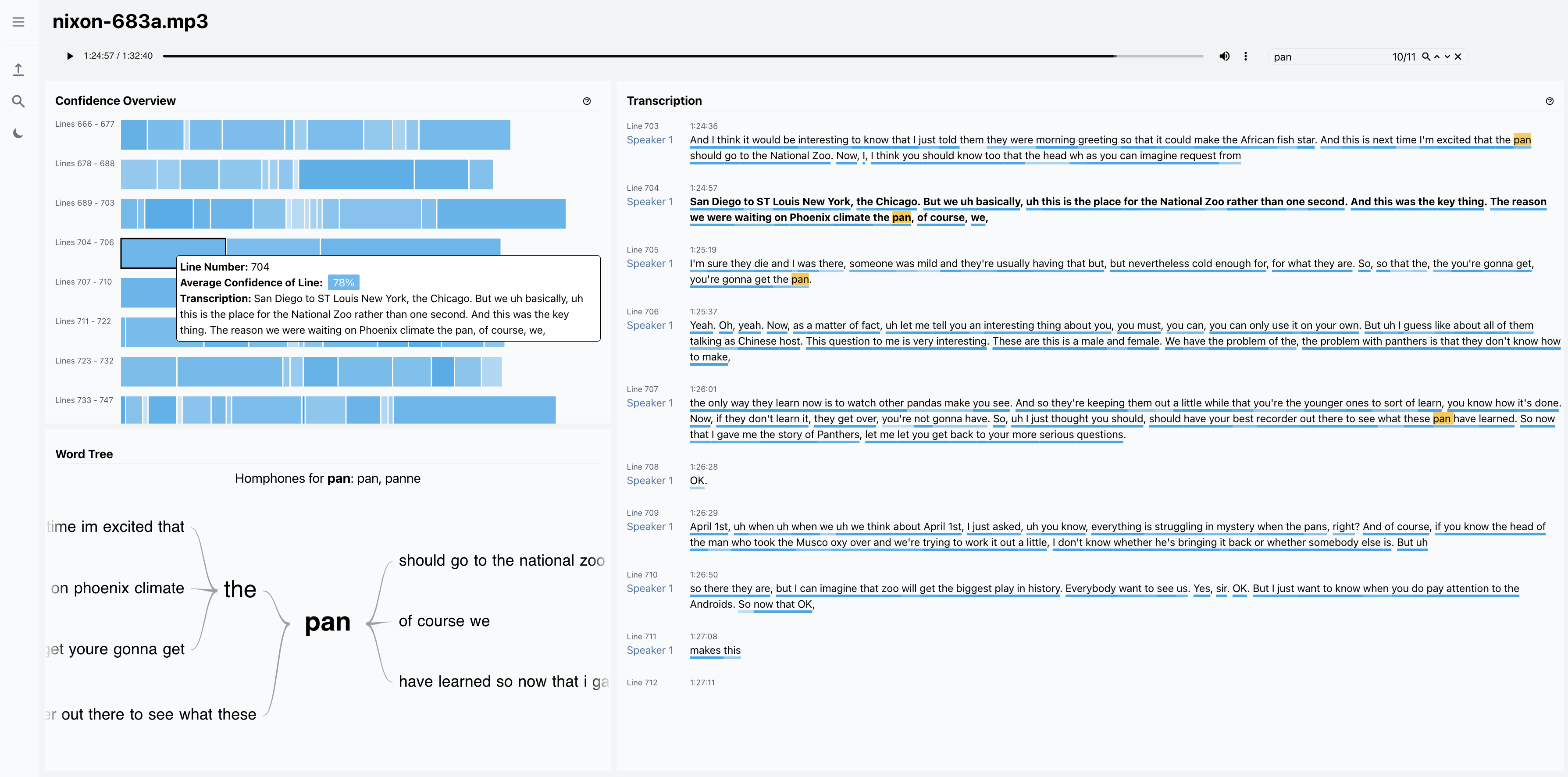}
    \caption{After searching for ``pan," we observe ``zoo'' in the word tree. This indicates that ``panda" was misclassified as ``pan" and hints that a zoo may be where the pandas will be kept.}
    \label{fig:case_study2}
    \vspace{-1em}
\end{figure}

\paragraph{\hspace{-3mm}\textbf{Which locations were considered for housing the pandas?} \href{https://youtu.be/hbeDn5D-GCg?si=DZ5w2Ojq3AfjIYqz&t=187}{\linkicon}}
As the analyst searches for more instances of ``pan," (See Figure~\ref{fig:case_study2}), they observe that word tree mentions ``national zoo.'' Additionally, line 704 lists several cities (San Diego, St. Louis, New York, and Chicago) known for their zoos. The transcription shows high confidence for the listed cities (ranging from 93\% to 100\%). Even though the confidence level for this particular segment is relatively high, the analyst decides to cross-check the source audio due to grammatical errors in the transcription. They discovered that despite considering the four listed cities, the pandas will be kept at the National Zoo in Washington, D.C., as one speaker reasoned that it is a tradition that all animals gifted to the U.S. are homed to this zoo. This can be uncovered by searching for ``zoo" and observing the first instance of the term, shown by line 631 in Figure~\ref{fig:teaser}.

\section{Discussion and Future Work}

Through this work, we aim to underscore the importance of AI transparency in fostering and calibrating appropriate trust in visual analytics, especially in scenarios involving domain experts making high-risk decisions. 
Research has shown that domain experts tend to perform tasks on their own~\cite{dasgupta_familiarity_2017,ha2024guided} despite the presence of AI assistance. To promote effective usage of AI tools, designers should carefully iterate through intuitive visual representations of uncertainty that do not add cognitive load to the analyst when performing a task~\cite{zhou2017effects}. However, even with the communication of model confidence, analysts are likely to engage with AI suggestions differently~\cite{sheridan_individual_2019}, influenced by their interpretation of confidence and their criteria for reliability~\cite{ottley2022adaptive}. Further research is needed to explore and develop design guidelines for visual presentations of confidence that are responsive to individual analysts’ needs within decision-making tools. 

In the transcription view of our system, we utilized underlining and color opacity to highlight uncertainty and possible errors within the output. As prior works have found inconclusive findings on the effectiveness of highlighting potential errors to aid in the correction of automatic speech-to-text data~\cite{vertanen_benefits_2008, suhm_multimodal_2001, soe_evaluating_2021}, future work involves conducting user studies to validate the system's ability to assist analysts in data cleaning and also uncovering relevant data more efficiently. We could also provide additional assistance to analysts through guided suggestions~\cite{ha_unified_2023} of relevant data points of interest for enhanced real-time exploration and analysis.
Currently, \system\ depends on AWS Transcribe to obtain automatic transcriptions and confidence values. Future directions can include decoupling from AWS Transcribe to leverage other services.

\section{Conclusion}
We developed \system\ alongside intelligence analysts to enhance the analyst's understanding of speech-to-text results by providing visual representations of uncertainty in several views of our system. In this paper, we identify clear design goals that guided the development of this system and explore a realistic use case scenario investigating the Nixon White House tapes to demonstrate the system's applicability. Finally, we discuss lessons learned from the development of this system and opportunities for improving textual data cleaning, promoting transparency, and fostering trust. Moving forward, continued efforts to improve the communication of model uncertainty and build appropriate reliance will further enhance the efficacy of human-machine collaborations within visual analytics.